# Maximizing the Bandwidth Efficiency of the CMS Tracker Analog Optical Links

Stefanos Dris[1,2], Markus Axer[1], Costas Foudas[2], Karl Gill[1], Robert Grabit[1], Jan Troska[1], Francois Vasey[1]

[1]CERN, 1211 Geneva 23, Switzerland
[2]Blackett Laboratory, Imperial College, London, UK
Stefanos.Dris@cern.ch

*Abstract*

The feasibility of achieving faster data transmission using advanced digital modulation techniques over the current CMS Tracker analog optical link is explored. The spectral efficiency of Quadrature Amplitude Modulation - Orthogonal Frequency Division Multiplexing (QAM-OFDM) makes it an attractive option for a future implementation of the readout link. An analytical method for estimating the data-rate that can be achieved using OFDM over the current optical links is described and the first theoretical results are presented.

## I. Introduction

The ~10 million channels in the Compact Muon Solenoid (CMS) Tracker sub-detector are read-out by 40 000 analog optical links [1]. These are embedded in the data acquisition system shown in Figure 1. The current optical links employ Pulse Amplitude Modulation (PAM) at 40MS/s. The Signal to Noise Ratio (SNR) of the system is specified so that the link has an equivalent digital resolution of at least 8 bits. Hence, the analog modulation scheme is akin to digital baseband PAM using 256 distinct levels (8 bits) at 40MHz. The equivalent data-rate is 320Mbits/s (=8×40MHz).

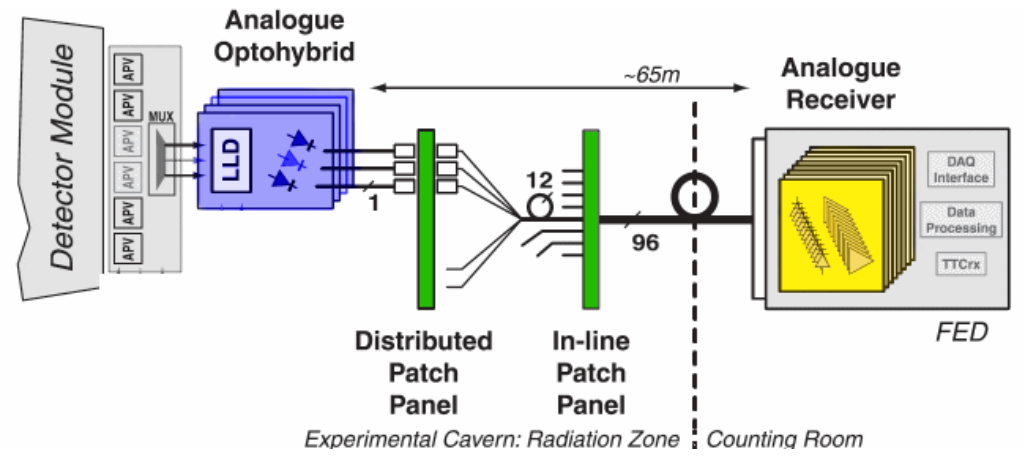

**Figure 1:** CMS Tracker readout system.

In the context of a SLHC upgrade [2], it is proposed to convert these links to a digital system in order to achieve higher data-rates using the current analog link components. The high cost of development of new optoelectronic components able to match the physical and environmental constraints of a high energy physics experiment provides the motivation behind re-using the current link components. Additional components on either side of the existing links would be required to perform the necessary digitization, digital transmission and reception. The main constraints are the bandwidth of the link and available signal power, limited by the transmitter hybrid (AOH) and the optoelectronic receiver amplifier (ARx12) [1]. Therefore, a bandwidth efficient digital modulation scheme is required to achieve transmission at Gbit/s rates.

The aim of this project is to determine the approximate data-rate and corresponding Bit Error Rate (BER) at which information could be transmitted using digital modulation over the current CMS Tracker analog optical links. Having identified the appropriate modulation scheme, the requirements in terms of hardware need to be laid out.

Basic concepts of communication theory and OFDM are introduced in Section II and the first part of section III of this paper. Section III.A is a detailed analysis of the data-rate calculation for a multi-carrier system, with references to texts describing underlying principles not covered in this paper. This is followed by the results (III.B) and conclusions.

## II. Digital Communication Basics

The work described in this paper refers to the development of a digital communication system, which can be represented by the generic blocks shown in Figure 2. In our case, the analog channel is the current CMS Tracker optical link. The choice of digital modulation (performed by the channel encoder block of Figure 2) depends on the characteristics of the channel through which signals are sent, and the trade-off between bandwidth-efficiency and transmitted power. This paper is primarily concerned with the channel encoding and analog channel blocks.

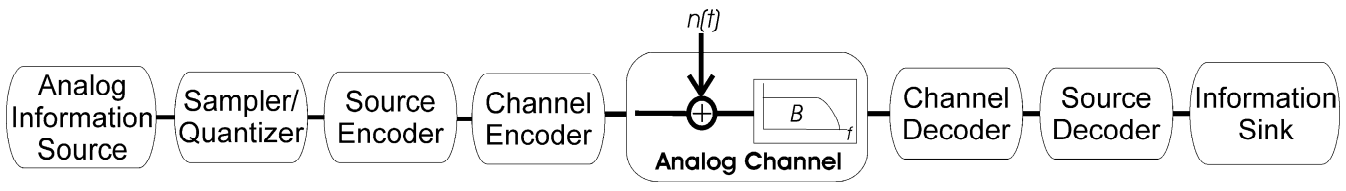

**Figure 2:** A Digital Communication System

### A. *The Channel: Current Optical Link*

The frequency response of a complete optical link was measured using a Spectrum Analyzer (Figure 3). A circuit based on a differential driver IC was used at the spectrum analyzer output in order to adapt it to the AOH input. The response of this circuit was subtracted from the response of the whole link. The ripple observed at higher frequencies could be a feature of the measurement system used, and are not necessarily a true representation of the link's response.

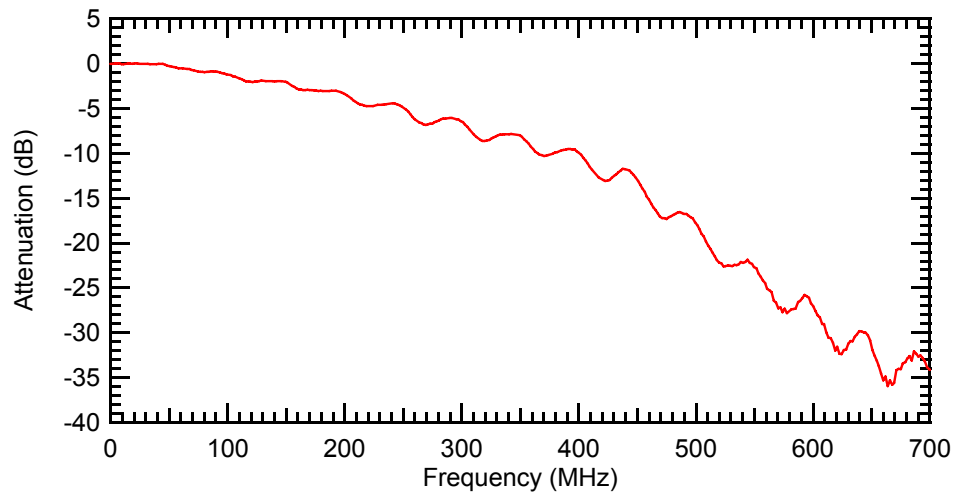

**Figure 3:** Frequency response of the analog link. The y-axis is the normalized signal power attenuation.

The noise of the link has been specified over 100MHz for a peak Signal to Noise Ratio (SNR) of 48dB [3]. In the analysis of the following sections, we will extrapolate and assume a constant noise power spectral density (PSD) for all frequencies.

## B. Channel Encoding: Bandwidth Efficiency

The encoder modulates the analog signal passing through the channel with digital information. Our objective is to maximize the data rate in our system, and hence bandwidth-efficient digital modulation schemes need to be considered.

In 1948, Claude E. Shannon defined the capacity, $C$, as the maximum number of information bits per channel use that can be transmitted over an additive white Gaussian noise (AWGN) communication channel with an arbitrarily small probability of error [4]:

$$C = \frac{1}{2} \cdot \log_2 (1 + SNR) \qquad \text{[bits/symbol]} \qquad (1)$$

The SNR in the above equation is defined as the ratio of average signal power divided by average noise power. The 'Shannon Limit' is useful for evaluating the performance of digital communication systems, since it is an objective metric; Shannon never specified what type of modulation, encoding or Forward Error Correction (FEC) is required to attain this limit. An alternative form of the Shannon Capacity equation can be written in terms of the channel bandwidth[1], $BW$:

$$C = BW \cdot \log_2 (1 + SNR) \qquad \text{[bits/s]} \qquad (2)$$

## C. Quadrature Amplitude Modulation

Quadrature Amplitude Modulation (QAM) is a bandwidth-efficient, passband modulation technique. The data in a QAM system is modulated onto a higher frequency sinusoidal carrier. Information is conveyed by varying both the amplitude and phase of the carrier. More details on QAM can be found in [5].

A single QAM carrier could be used in the CMS Tracker optical link to increase the bandwidth efficiency compared to the current PAM scheme. However, QAM requires the channel to have a flat magnitude response over the bandwidth it occupies. Our channel is frequency-selective and this causes InterSymbol Interference in the QAM signal [6]. A typical solution would be to employ an equalizer at the receiver. This is a digital filter with a frequency response that is the inverse of that of the channel. At present, the variation in frequency response from link to link is not known, meaning we cannot be sure that one set of equalizer coefficients would be suitable for all optical links. If this is not the case, a computationally expensive adaptive equalizer would have to be designed for the receiver, and this is probably impractical for the Gbit/s data-rates that we hope to achieve. Nevertheless, single-carrier modulation should not be ruled out, and this option will be explored in the near future.

[1] The term 'bandwidth' refers to the full frequency range of the channel.

## D. Multiple QAM Carriers

An alternative way of avoiding ISI in a frequency-selective environment is to divide the channel into $N$ sub-channels, and place a QAM modulated carrier in each. For a large enough $N$, each sub-channel will have a flat frequency response, hence eliminating the need for equalization at the receiver. Figure 4 is an illustration of the sub-channel concept, with each arrow representing a QAM modulated carrier.

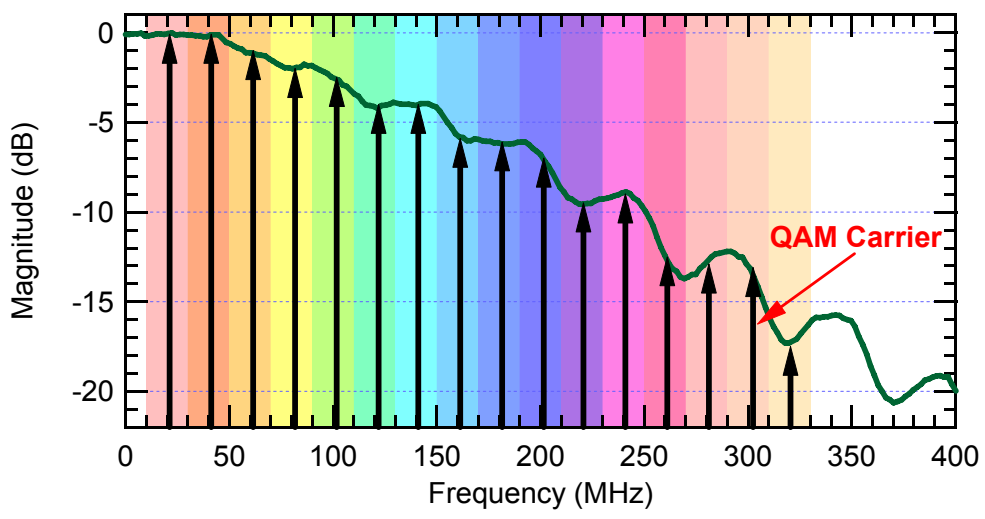

**Figure 4:** Illustration of a multiple carrier system

# III. Orthogonal Frequency Division Multiplexing

A multi-carrier scheme which is of great interest is Orthogonal Frequency Division Multiplexing (OFDM) [5] [6]. It is widely used in high-data rate applications such as ADSL and Wi-Fi. OFDM makes efficient use of the channel bandwidth by allowing spectral overlap between adjacent carriers. For reception without InterCarrier Interference (ICI) the carriers must be mathematically orthogonal. The receiver acts as a bank of demodulators, translating each carrier down to DC, with the resulting signal integrated over a symbol period to recover the baseband data. If the other carriers have frequencies that, in the time domain, have an integer number of cycles in the symbol period $T$, then the integration process results in zero contribution from all these other carriers. Thus, the carriers are linearly independent if the carrier spacing is a multiple of $1/T$. This is illustrated in Figure 5.

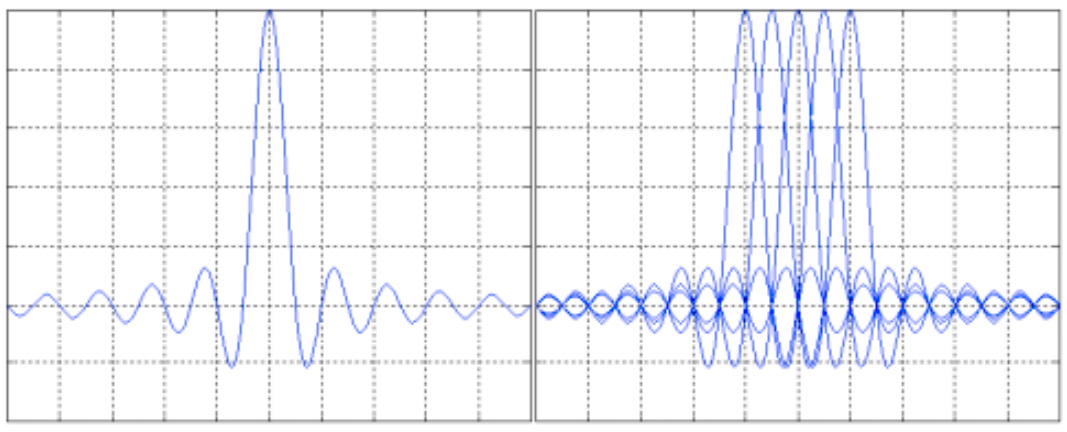

**Figure 5:** Spectra of a single (left) and four overlapping (right) OFDM sub-channels

## A. OFDM System Design

In this section we attempt to estimate the data-rate possible using OFDM over the CMS Tracker optical link. The analysis is based mainly on the work by Cioffi [7] [8]. We start by discussing the peak power issues in OFDM. We then introduce the concept of the SNR gap [8] that determines the number of bits/symbol (and therefore data-rate) for a single QAM carrier. Finally we define the multi-channel SNR [8], a measure that defines an equivalent single carrier performance metric for a multi-carrier system, and this simplifies the calculation of the achievable data-rate.

*1) OFDM Power Issues*

A major drawback of OFDM is the high peak-to-average power ratio (PAPR) of the transmitted signal [9]. The peak signal occurs when all sub-carriers add coherently, and it can be shown that the peak power of an OFDM signal is equal to the number of carriers ($N$) times the average power. Clearly, PAPR is a problem for the transmitter amplifier, since it must accommodate these large signals in order to avoid signal clipping which leads to intermodulation interference and therefore bit-errors. This is achieved by lowering (or 'backing off') the average power of the transmitted signal, so that the signal peaks fit in the transmitter's input range. Power consumption of a power amplifier depends largely on the peak power, and hence accommodating occasional large peaks leads to low power efficiency. This so-called power 'backoff' leads to lower signal power, and hence lower SNR available for data transmission.

As an example, consider the case of a multi-carrier system with a PAPR=256. The backoff required for completely avoiding signal clipping is $10\cdot\log_{10}(256)=12$dB. However, the resulting SNR reduction could lead to a severely lower data-rate for a given BER. Practical OFDM systems do not employ such a large backoff, at the expense of a tolerable degradation in BER. The chosen backoff of the transmitted signal is thus a compromise between the contradictory desires to achieve a large average transmit power on the one hand and a low distortion due to clipping of the signal on the other [9].

However, high signal peaks in OFDM occur rarely. The amplitude of an $N$-carrier OFDM signal has a Rayleigh distribution with zero mean, and a variance of $N$ times the variance of one complex sinusoid [10]. Given the signal statistics, one can calculate the BER penalty incurred due to signal clipping and select the optimum power backoff for a particular implemented system. This calculation is implementation-specific, and is beyond the scope of this work. We will, instead, use typical backoff values from the literature [11] [12] in our calculations, and show what data-rate can be attained, ignoring any resultant BER degradation.

*2) SNR Gap Analysis*

'SNR gap analysis' will be described here for a single sub-channel (i.e. one QAM carrier) as an introduction to the concepts relevant to multi-carrier system design. The equations needed for calculating the maximum bit-rate in a single-carrier QAM system are presented.

A channel with double-sided noise power spectral density $N_0$, signal energy $\mathcal{E}$, and channel gain $|H|$ has a $\text{SNR}=\mathcal{E}\cdot|H|^2/N_0$. The maximum bit-rate achievable over a bandwidth, $BW$, in this channel is given by a variation of the Shannon Capacity:

$$b = BW \cdot \log_2\left(1+\frac{SNR}{\Gamma}\right) \qquad (3)$$

$\Gamma$ is the 'SNR gap'. When $\Gamma=1$ (0dB), Eq.3 is equal to Eq.2, and $b$ is the capacity of the channel. Any reliable and implementable system must transmit at a rate below capacity, and hence $\Gamma$ is a measure of loss with respect to theoretically optimum performance.

The BER probability in QAM is upper-bounded by the symbol error probability, $P_e$, which is closely approximated by [6]:

$$P_e \leq 4Q\left(\sqrt{\frac{d_{\min}^2}{2N_0}}\right) \qquad (4)$$

The Q-function in the above equation is given by:

$$Q(x) = \frac{1}{\sqrt{2\pi}}\int_x^{\infty} e^{x^2/2}dx \qquad (5)$$

$d_{min}$ is the minimum distance between QAM constellation points at the channel output:

$$d_{\min}^2 = d^2|H|^2 \qquad (6)$$

$|H|$ is the channel gain and $d$ is the distance between points in the uncoded input constellation. It is clear that the BER depends on the noise PSD and $d_{min}$, itself depending on the signal power (more allocated power means greater distance between the symbols in the constellation and hence larger SNR). The SNR Gap, $\Gamma$ (sometimes called the 'normalized SNR') is defined by [7]:

$$3\cdot\Gamma = \frac{d_{\min}^2}{2N_0} \qquad (7)$$

Note that if we take the square root of $3\cdot\Gamma$ we obtain the argument to the Q-function of Eq.4. Hence, we can calculate $\Gamma$ by first deciding on a target BER, calculating the corresponding Q-function argument from Eq.4, and substituting into Eq.7. Substituting $\Gamma$ into Eq.3 then gives the achievable data-rate of a single-carrier QAM system.

*3) Multi-Carrier QAM Data-Rate*

The analysis of the previous section can be extended to OFDM by recognizing that the multiple carriers occupying sub-channels are independent to each other. Hence the aggregate performance of the individual QAM carriers in their sub-channels gives the performance of the multiple carrier system in the complete channel.

The total BER in an OFDM system is simply the average of the individual sub-carrier BERs. We assume that all carriers carry equally important information, and hence we require the same BER in each. Given the SNR in each sub-channel, we can then determine the individual carrier bit-rates as described in the previous section. The total bit rate is found by simple addition. In equation form:

$$b_{tot} = BW_k \cdot \sum_{k=1}^{N}\log_2\left(1+\frac{SNR_k}{\Gamma}\right) \qquad (8)$$

$BW_k$ is the channel spacing (equal to the symbol rate for each of the $N$ sub-channels), while $SNR_k$ denotes the SNR for the $k_{th}$ carrier.

*4) Multi-carrier Power Allocation*

There is a subtle point about the SNR per sub-channel that needs to be understood before proceeding. The SNR in each OFDM sub-channel depends on how the total power budget is allocated. Since we are assuming the noise is white (constant PSD, $N_0$ W/Hz across the whole bandwidth of the channel), the SNR of any given sub-channel will then depend on the power allocated to the corresponding carrier.

The easiest approach would be to allocate equal power to all carriers. Let $N$=number of carriers, and $\mathcal{E}_{total}$=total power budget. The power per carrier would then be $\mathcal{E}_{total}/N$. However, this is a non-optimal solution due to the fact that we must impose the restriction that all QAM constellations used in our uncoded OFDM system must have an integer number of allocated bits. It can be easily seen that Eq.3 can yield a fractional bit allocation for a particular carrier – say 5.3 bits/symbol. If we are to meet our target BER, we would have to round down to 5 bits/symbol. This would be a 'waste', since part of the power allocated would effectively be used for lowering the BER in that particular sub-channel. It would be more sensible to allocate this superfluous power to another carrier in order to increase its SNR and hence its number of allocated bits.

Moreover, there are limitations imposed by the QAM constellation 'shapes'. Ideally, we want rectangular constellations (i.e. ones where the symbols can be evenly spaced, forming a rectangle). For example, if the SNR in a sub-channel allows transmission of 3 bits/symbol, there is no efficient way of arranging $2^3$ symbols in a QAM constellation. In such a case, it would be preferable to allocate one less or one more bit to this sub-channel. This would have to be done by adjusting the power allocated to that particular carrier.

The optimum power and bit-allocation is achieved using the so-called 'water-filling' optimization [8]. The exact description of the algorithm is beyond the scope of this document, but a formulation of the general optimization problem is given here. Our goal is to maximize the sum given in Eq.8 by varying the allocated bits/symbol and power. Let $g_k$ be the sub-channel SNR when unit energy is applied to the $k_{th}$ carrier:

$$g_k = \frac{|H_k|^2}{N_0} \qquad (9)$$

Substituting Eq.9 into Eq.8 we have:

$$b_{tot} = BW_k \cdot \sum_{k=1}^{N} \log_2\left(1 + \frac{\mathcal{E}_k \cdot g_k}{\Gamma}\right) \qquad (10)$$

$g_k$ is a constant for the channel, but $\mathcal{E}_k$ can be modified to maximize $b_{tot}$, subject to the constraint of the total power budget:

$$\sum_{k=1}^{N} \mathcal{E}_k = \mathcal{E}_{total} \qquad (11)$$

*5) Equivalent Single-carrier Metric: Multi-channel SNR*

We can, in fact, ignore the effects of the bit and power allocation algorithm in determining the upper bound on the achievable data-rate in a multi-carrier system. The following analysis makes the calculation extremely straightforward, though one must not forget that the actual data-rate achieved in a real system will depend on the chosen power allocation.

We start by assuming a constant BER (and SNR gap) across all sub-channels. This allows for the use of a single performance measure to characterize the multi-channel transmission system. This measure is a geometric SNR ($SNR_{m,u}$) and can be compared directly to the detection SNR of a single-carrier system employing equalization.

For a set of $N$ parallel channels of symbol rate $BW_k$ each, the bit rate is:

$$\begin{aligned} b &= \sum_{k=1}^{N} BW_k \cdot \log_2\left(1 + \frac{SNR_k}{\Gamma}\right) \\ &= BW_k \log_2\left(\prod_{k=1}^{N}\left[1 + \frac{SNR_k}{\Gamma}\right]\right) \qquad (12) \\ &= N \cdot BW_k \cdot \log_2\left(1 + \frac{SNR_{m,u}}{\Gamma}\right) \end{aligned}$$

$SNR_{m,u}$, the multi-channel SNR for a set of parallel sub-channels is defined by:

$$SNR_{m,u} = \left[\left(\prod_{k=1}^{N}\left[1 + \frac{SNR_k}{\Gamma}\right]\right)^{1/N} - 1\right] \cdot \Gamma \qquad (13)$$

The multi-channel SNR characterizes all sub-channels by an equivalent single AWGN channel that achieves the same data-rate. In our calculation we use Eq.12 and Eq.13 to estimate the achievable data-rate of an OFDM system given the noise PSD and measured frequency response of our channel.

*6) Approximate Data-rate Calculation Using Multi-channel SNR*

Let the channel be divided into a large number of sub-channels, say $N$=256, with equal power allocated to each sub-channel. For this approximate solution to work, we need to determine the total frequency range that will be occupied by our QAM carriers. The low-pass response of our channel makes this straightforward, since we need only find the highest frequency at which the $256^{th}$ carrier can transmit 1bit/symbol. Since we are assuming equal power distribution in all carriers, it follows that they must all transmit data. Modifying Eq.3 to a more general form, we can calculate the number of bits/symbol on a QAM carrier, given a channel SNR:

$$C = \frac{1}{2} \cdot \log_2 \left( 1 + \frac{SNR}{\Gamma} \right) \quad (14)$$

Rearranging:

$$SNR = \Gamma \cdot (2^{2C} - 1) \quad (15)$$

By setting C=1 (and with Γ determined by the target BER using Eq.4 and Eq.7), the minimum SNR required for transmitting 1 bit/symbol can be found. As an example, consider the case where a target BER of $10^{-9}$ is set. The required SNR for transmitting 1bit/symbol is then ~16dB. A normalized (with respect to a maximum input signal of 1W) value of $N_0$ has been calculated for our channel, and the power allocated to each carrier in our calculation can be found from $\mathcal{E}_k = \mathcal{E}_{total}/256$. The SNR at the $k^{th}$ sub-channel is then given by:

$$SNR_k = \frac{P_k \cdot |H|^2}{BW_k \cdot N_0} \quad (16)$$

When $k$=256, $P_k$ and $BW_k$ denote the allocated power and bandwidth of the 256th (last) sub-channel. In OFDM, the sub-channel bandwidth (i.e., carrier spacing) is determined by the chosen symbol rate, and hence, we need to find the carrier frequency at which the $SNR_{256}$~16dB. This frequency (we shall call it $f_{max}$) is the upper limit to the channel bandwidth that can be used.

From the above, we have also determined the symbol rate per carrier of our theoretical system (=$f_{max}$/256). Hence, the 256 QAM carriers will be centered at the following frequencies:

$$f_k = k \cdot symbol\ rate \qquad for\ k = 1\ to\ 256 \quad (17)$$

From Eq.16, we calculate the SNR at each sub-channel. Substituting these values into Eq.13, and given the value of the SNR gap, we obtain $SNR_{m,u}$. Finally, substituting into Eq.12, we determine the bit-rate of our system.

## *B. Results*

The method described in the previous section was used to estimate the achievable data-rate using OFDM over the CMS Tracker Analog optical links. Of course, the accuracy of the results depends on the accuracy of the assumptions. In order to produce an unbiased first-order estimate and evaluate the effect of errors, we have varied the main design parameters to obtain a range of data-rates. Firstly, various target error rates were used. While Eq.4 is only an approximation of the upper bound on the BER, this uncertainty is effectively removed when changing the error rate by a few orders of magnitude.

The total power available to the system was estimated given the linear input range to the AOH (transmitter). The peak signal that can be sent through the link is assumed to be 600mV. Non-linearities within and outside this input range are not considered in the analysis. In our analysis the peak power has been normalized to 1W, and this corresponds to using the full AOH input range. We have used a number of power backoff values, knowing that typical numbers found in literature regardless of carrier number are 3-9dB [11] [12].

Finally, the noise PSD is an estimate based on specifications and is assumed to be constant across all frequencies. This probably over-estimates the noise, and hence it is the cautious approach. A range of constant noise PSD values was used in order to determine the effect that a wrong estimate can have on the achievable data-rate.

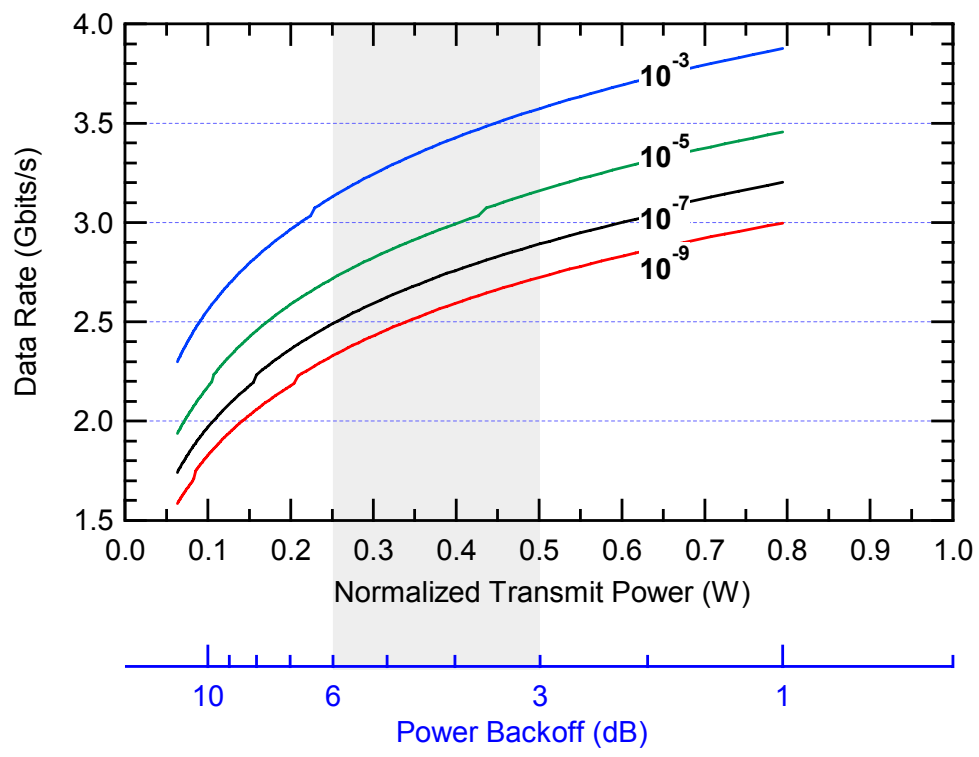

**Figure 6:** Data-rate as a function of transmitted signal power (and power backoff in dB) for various target BERs.

Figure 6 shows the data-rate against normalized transmitted signal power (and power backoff in dB). The results are shown for four different target BERs. The shaded area corresponds to the backoff range of 3-6dB, which is the range we are likely to operate in. At a BER of $10^{-9}$, the data rate in this region of operation ranges from 2.3 to 2.7Gbits/s. The corresponding bandwidth efficiency is ~6-6.5bits/s/Hz. This can be compared to the bandwidth efficiencies of Wi-Fi (~3bits/s/Hz), and ADSL (typically up to ~6bits/s/Hz).

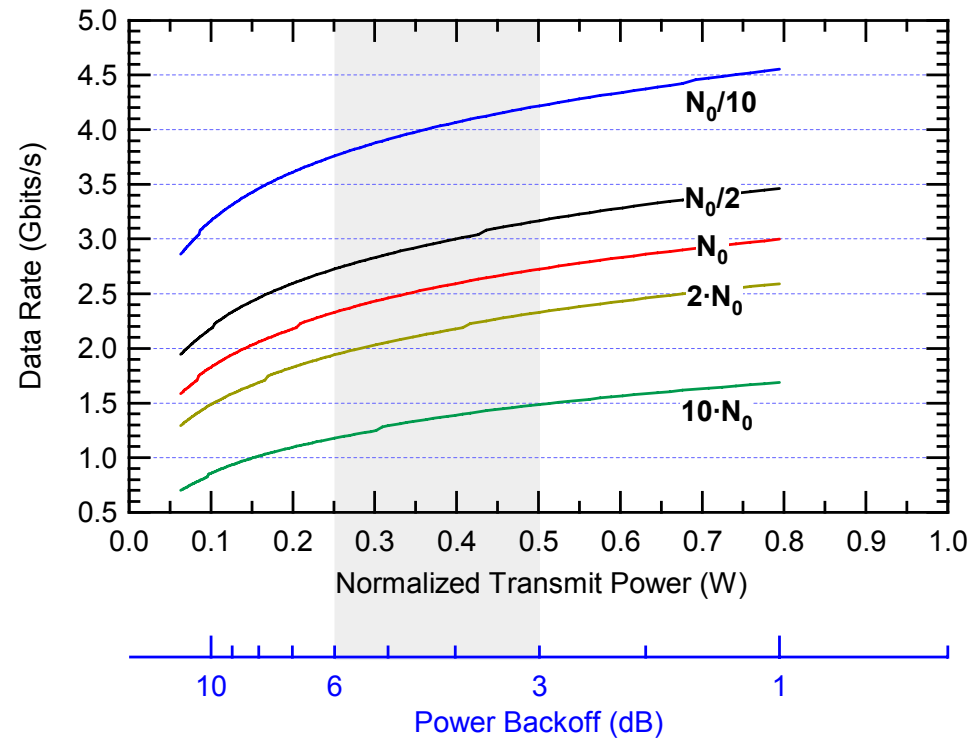

**Figure 7:** Data-rate as a function of transmitted signal power (and power backoff in dB) for various noise PSDs.

In order to assess the impact of the possible error in the average noise PSD used in the calculations, the data-rate for a fixed error rate (BER=$10^{-9}$) was computed as a function of transmitted signal power for various noise PSDs. The results are shown in Figure 7. The middle curve corresponds to the noise PSD, $N_0$, calculated from the specification of the optical link. Results were obtained for higher noise ($10 \cdot N_0$ and $2 \cdot N_0$) as well as lower noise ($N_0/10$ and $N_0/2$). The result shows that if the error on the average noise PSD is of the order of ±100%, the calculated data-rate varies by ±400Kbits/s (±15%).

In practice, the data-rate in an OFDM scheme could be lower by ~10-20% due to the inclusion of the so-called cyclic prefix, which is essentially a sequence of redundant information. This is required to maintain orthogonality of the carriers, and is normally chosen to be at least as long as the impulse response of the channel [5].

The above calculations do not include all effects that may be encountered in a practical system. Phase noise, non-linearities in the link transfer function and frequency instabilities can adversely affect the system's performance. At this stage, we have little or no data that can be used to evaluate the significance of these effects on our chosen modulation scheme. It follows that performance of any advanced digital communication scheme is heavily dependent on the actual hardware implementation.

Another source of uncertainty is the frequency response measurement made on the optical link, due to the limited accuracy of the instrument used, as well as the presence of components that would not be present in a future implementation; a differential driver circuit was used before the AOH, while the ARx12 sat on a board that included output buffer amplifiers.

Finally, it should be noted that the results are for an uncoded system. Forward Error Correction (FEC) and other codes can be used to decrease the BER in the system, hence allowing transmission at higher speeds. These will be investigated in the future.

### *C. Future Plans*

In order to assess the accuracy of the assumptions made in this paper, we intend to conduct lab measurements involving the transmission of QAM signals through the link. A single-carrier QAM signal will be created using an Arbitrary Waveform Generator desktop instrument interfaced to the AOH. The signal will be demodulated either by a Spectrum Analyzer, or captured by a Real-Time scope and demodulated offline using MATLAB. This test will ultimately determine the feasibility of using QAM through the optical link. It will be possible to assess the impact of amplitude and phase noise, and hence determine the number of bits/symbol that can be transmitted on a QAM carrier at various frequencies. One can also establish the maximum symbol rate that can be used before ISI becomes an issue. The knowledge gained will then be used to extrapolate the performance of a multi-carrier scheme.

## IV. Conclusions

We have described the first attempt to calculate the achievable data-rate using a bandwidth-efficient modulation scheme over the CMS Tracker Analog optical link. The calculation is based on test data and specifications of current link components, and where necessary, extrapolations have been made to account for unknown characteristics of the link.

Based on the assumptions made, data-rates of over ~2.5 Gbits/s should be possible in a future system based on the current link components. Hardware verification tests are planned, and the next step will be to assess the complexity of an OFDM implementation. It will then be possible to evaluate whether the data-rate increase is worth the investment in extra components and implementation complexity.

## V. Acknowledgements

The authors would like to thank Dr Izzat Darwazeh, (Reader in Telecommunications Engineering, University College London), Dr Ersi Chorti (Communications and Signal Processing, Imperial College London), Dr Nicholas Avlonitis (Optical and Semiconductor Devices, Imperial College London) and Professor Eric Yeatman (Optical and Semiconductor Devices, Imperial College London) for the discussions and valuable comments.